# Input Impedance, Nanocircuit Loading, and Radiation Tuning of Optical Nanoantennas

Andrea Alù, and Nader Engheta[*]

Dept. of Electrical and Systems Engineering

University of Pennsylvania

Philadelphia, PA 19104, U.S.A.

**Abstract**

Here we explore the radiation features of optical nanoantennas, analyzing the concepts of input impedance, optical radiation resistance, impedance matching and loading of plasmonic nanodipoles. We discuss how the concept of antenna impedance may be applied to optical frequencies, and how its quantity may be properly defined and evaluated. We exploit these concepts in optimization of nanoantenna loading by optical nanocircuit elements, extending classic concepts of radio-frequency antenna theory to the visible regime for the proper design and matching of plasmonic nanoantennas.

---

[*] To whom correspondence should be addressed.  E-mail: engheta@ee.upenn.edu

The anomalous electromagnetic features of plasmonic materials and components at optical frequencies have been recently applied to various fields, and in particular to the realization of optical devices with reduced size and enhanced functionalities The research is currently leading towards the realization of integrated plasmonic devices that may have functionalities similar to their radio-frequency (RF) counterparts, but with much enhanced bandwidth, speed and compactness. This interest has been accompanied by remarkable advancements in nanotechnology, fabrication and measurement of nanostructures with the desired geometry and electromagnetic properties in the nanometer scales.

Driven by these advancements, and in particular by the possibility of realizing plasmonic nanorods with elongated shapes and relatively high aspect ratios, several groups have experimentally investigated the realization of nanoantennas in the form of monopole and dipole antennas made of plasmonic materials [1]-[6]. From a theoretical point of view, however, the recipe for proper design and performance optimization of these nanoradiators is still in its embryonic stage when compared with well-known design methods for conventional RF antennas. A recent letter [7] has pointed out the shortening of the effective wavelength "seen" by the plasmonic dipole with respect to the free-space wavelength, consistent with the well known slow-wave properties of plasmonic nanorods [8]. All such studies, however, have been concerned with the scattering properties of these components when illuminated by external sources, e.g., light beams or microscope tips, or with their interaction with nanostructures nearby. It is interesting to note however that, by definition, one of the main functions of any antenna includes 'connecting' and 'matching' a source, a receiver, or a waveguide to the 'outside' region (e.g., free space domain) in order to transmit (or receive) a given signal to (or from) the far field. In this sense, the concepts of input impedance at the feeding point, antenna's radiation resistance, and the role of loading an antenna, all of which are familiar notions for traditional RF antennas, are quantities of fundamental importance for a proper design and use of antennas. These parameters, however, cannot be obtained from the scattering analysis. This is why, for instance, in [9] a simplified transmission-line analysis has been applied to the conduction current distribution over a carbon nanotube, deriving to first approximation its radiation properties, and pointing out some potentially important differences between RF and higher-frequency dipoles.

In this letter we transplant the concepts of input impedance, radiation resistance, loading and impedance matching into the optical domain, and we investigate them for plasmonic nanoantennas in the form of plasmonic nanodipoles. These results may lead to a paradigm for bringing the classical antenna concepts from RF to optical frequencies, in order to facilitate a realistic design and use of such nanodevices as efficient radiating or receiving optical elements at nanoscales for potential applications in telecommunications, super-resolution microscopy, biological and medical optical sensing.

The geometry of the problem is depicted in Fig. 1, which consists of a cylindrical nanodipole of total length $L$ and radius $R$, made from silver, and terminated on both sides by spherical tips. The dipole is assumed to be fed at its center gap of thickness $g$. This resembles a conventional center-fed RF dipole antenna;

however, here the optical nanoantenna is made of silver, a low-loss noble metal whose permittivity in the IR and optical regime may be described as a classic Drude model $\varepsilon_{Ag} = \varepsilon_0 \left[ \varepsilon_\infty - f_p^2 / \left( f \left( f + i\gamma \right) \right) \right]$ with $\varepsilon_\infty = 5$, $f_p = 2.175\,PHz$ and $\gamma = 4.35\,THz$ [10].

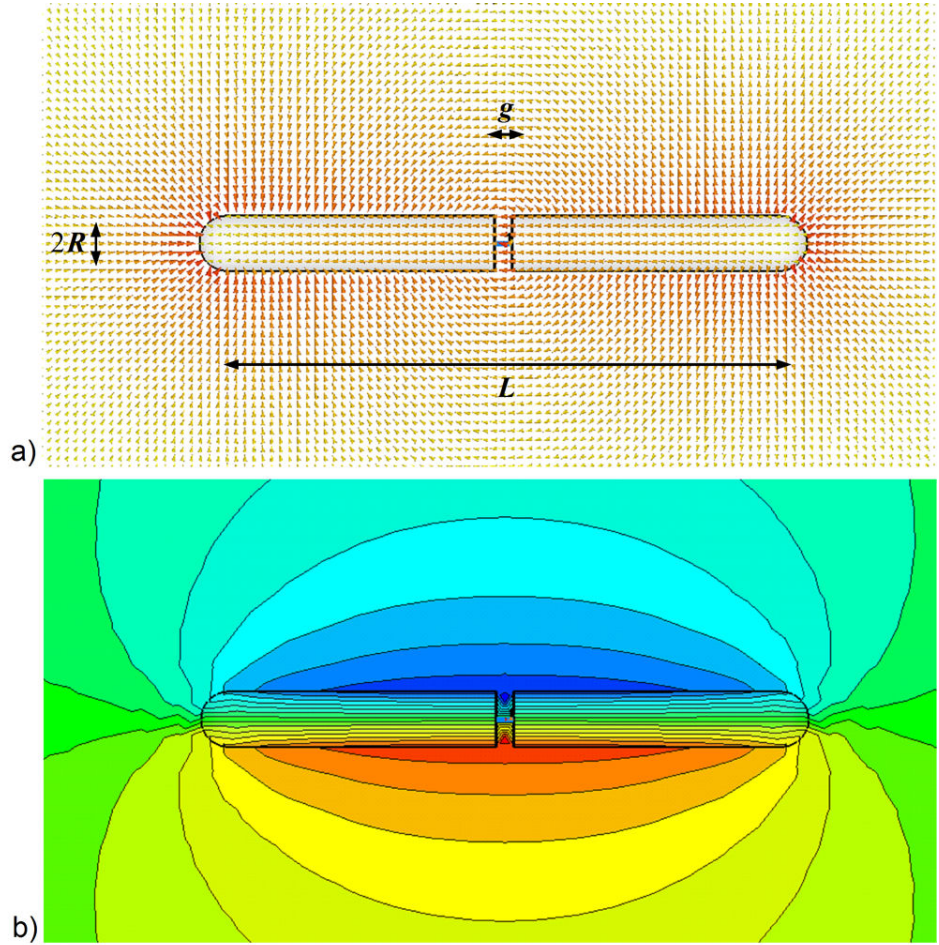

Figure 1 – (Color online) Geometry and (s) electric and (b) magnetic field distributions on the E plane (snapshots in time) at the resonant frequency for which $L \simeq \lambda_{eff}/2$ for a silver nanodipole antenna. In this case, $L = 110\,nm$ and $f = 266\,THz$. Brighter colors and longer arrows correspond to stronger values of the field.

Feeding the nanoantenna at its gap allows one to determine its input impedance $Z_{in} = R_{in} - iX_{in}$, defined as the ratio of the driving optical voltage difference across the gap to the total flux of induced optical displacement current flowing from the feeding source into the antenna (and the gap). Figs. 2a and 2b report $R_{in}$ (resistive component) and $X_{in}$ (reactive component) for different lengths of the nanodipole, while $R = 5\,nm$ and $g = 3\,nm$ kept fixed, using full-wave time-domain simulation software [11]. For very low frequencies, the silver is a conductive metal and the corresponding impedance is strongly capacitive (i.e., $X_{in} < 0$), analogous to a short RF dipole. However, increasing the frequency up to the IR and visible regime, the plasmonic features of silver come into play and the antenna hits its first resonance at the frequency for which $X_{in} = 0$, indicated, as an example, by the blue arrow in Fig. 2b for the $L = 160\,nm$ curve. It is evident that by decreasing the length $L$, the antenna resonant frequency shifts up, analogous to the case of a regular RF dipole. Its position is located around the frequency for which the condition $L = \lambda_{eff} / 2$ holds, where $\lambda_{eff}$ is the effective wavelength "seen" by the plasmonic dipole, which is shorter than the free-space wavelength $\lambda_0$, and it may be determined with good approximation as the wavelength of the guided $TM_{10}$ mode supported by an infinite silver nanorod of raidus $R$, yielding results consistent with the approximate linear formula derived in [7]. The value of the input resistance at this resonance frequency is lower than that of a regular RF dipole antenna (for instance $R_{in} = 22.2\,\Omega$ for the dipole of Fig. 1 with $L = 110\,nm$), as discussed in the following. Consistent with the transmission-line model of a dipole [12], this first resonance has a low input impedance, and it may be seen as a "short-circuit" resonance (i.e., $X_{in} = 0$) associated with the condition $L = \lambda_{eff}/2$. In contrast, a sharp "open-circuit" resonance (i.e., where $R_{in}$ and $X_{in}$ very large) is also noticed in Fig. 2 at a somewhat higher frequency, which corresponds to the sharp impedance resonance in the plots. The impedance values associated with this resonance here are larger than those obtained at the $L = \lambda_0$ resonance for a regular conducting RF dipole antenna with similar aspect ratio, owing to the plasmonic properties of the nanoantenna. Moreover, the "open-circuit" resonant frequency is lower than the $L = \lambda_{eff}$ frequency, due to the non-negligible capacitance of the air gap region. The input impedance evaluated in Fig. 2 may indeed be regarded as the parallel combination of the "intrinsic" impedance of the nanodipole $Z_{dip}$ and the air gap capacitive impedance $Z_{gap}$. This is consistent with the circuit model depicted in the inset of Fig. 2. This second resonance therefore arises when the inductive reactance of the nanodipole itself (at a frequency higher than the short-circuit resonance) "goes into resonance" with the air gap capacitance, resulting in large input impedance. The "open-circuit" resonance coincides with the scattering resonant frequency, when the unfed dipole is illuminated by an external plane wave, since the series combination of $Z_{dip}$ and $Z_{gap}$ resonates at this frequency.

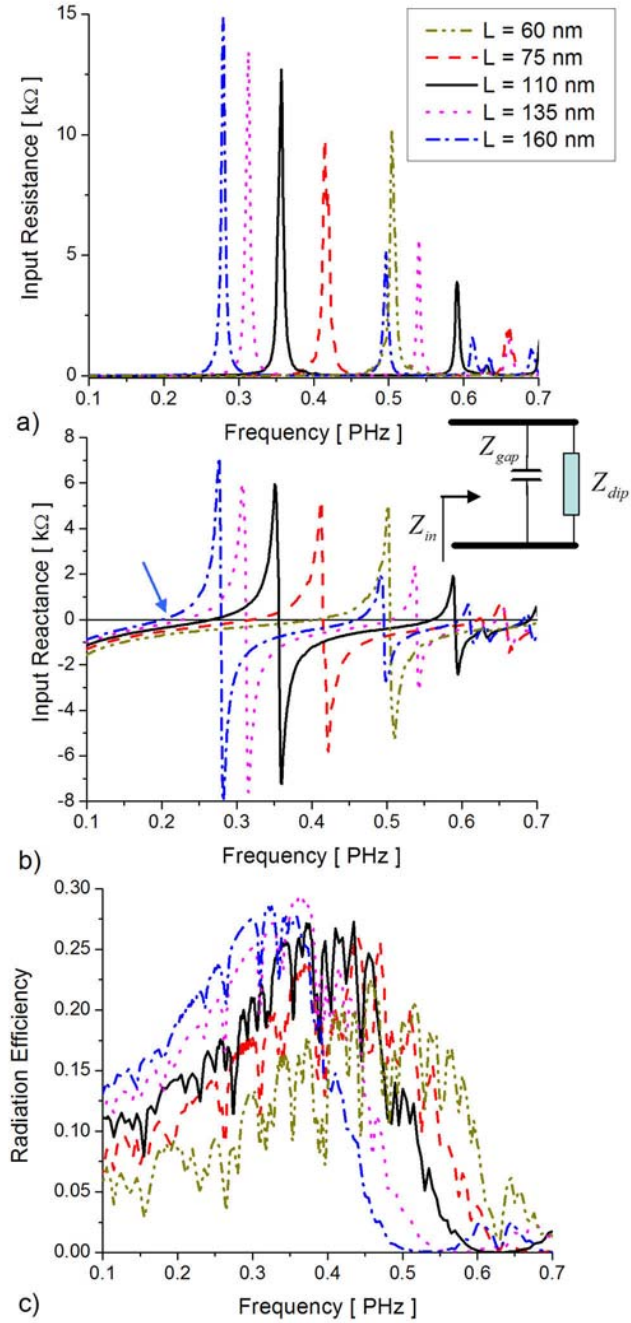

Figure 2 – (Color online) (a) Input resistance ($\text{Re}(Z_{in})$), (b) input reactance ($-\text{Im}(Z_{in})$) and (c) radiation efficiency for the dipole nanoantenna of Fig. 1 made of silver, with $R = 5\,nm$ and $g = 3\,nm$ varying $L$, calculated numerically using [11].

A regular RF radiating antenna is usually operated at the "short-circuit" resonance for matching reasons, but for the optical nanoantenna operation both these resonant frequencies may be appealing, depending on the impedance of the feeding mechanism (it may be noted that the peak values of input resistance and reactance in Figs. 2a, 2b around the "open-circuit" resonance may be comparable with the characteristic impedance of plasmonic waveguides [13]).

In Fig. 2c, we have evaluated the optical nanoantenna efficiency $\eta_{eff}$, known as the ratio between the total radiated power $P_{rad}$ and the power effectively "accepted" by the antenna input terminals. Although the peak of efficiency is located near the open-circuit resonance, this optical radiation efficiency is reasonably high over a wide frequency window around $L = \lambda_{eff}/2$, despite the material loss in silver and the nanodipole plasmonic features.

Figure 1 reports the electric and magnetic field distribution in the E plane for a nanodipole with $L = 110\,nm$ at the "short-circuit" frequency $f = 266\,THz$ (at this frequency $\lambda_{eff} \simeq 255\,nm$, slightly larger than the resonance condition $\lambda_{eff} = 2L$, due to the reactive fields at the end of the dipole, which may be approximately taken into account by introducing an effective dipole length $L_{eff} \simeq L + 2R$, consistent with [7]). It is noted that both field distributions resemble those of a conventional half-wavelength resonant RF dipole, ensuring reasonably good radiation and matching at the input port. However, we note the peculiar longitudinal electric field distribution inside the silver dipole, responsible for the flow of optical displacement current along the antenna. This current effectively plays the role of the conduction current flowing on the surface of an RF dipole made of highly conductive material.

The nanodipole antenna supports other higher-order resonances with different modal distributions around the frequencies for which $L \simeq N \lambda_{eff} / 2$, with $N$ being an integer (the shortening of $\lambda_{eff}$ with frequency is faster than linear, causing a denser increase of modes at higher frequencies). Due to their modal distributions, however, the higher-order resonant modes do not provide a significant far-field radiation, as confirmed by the low radiation efficiency (Fig. 2c) outside the range of frequency for which $\lambda_{eff} \simeq 2L$.

Analogous to the case of RF antennas made of highly conductive metal, for the plasmonic optical nanodipole here we may define an optical radiation resistance as $R_{rad} = P_{rad} / (2 I_{max}^2)$, with $I_{max}$ being the peak of current along the dipole. However, since silver behaves differently at optical frequencies, following our previous discussions in optical nanocircuits [14]-[15] and consistent with Fig. 1a, the current flow along the nanodipole should consider now the flux of optical displacement current, denoted as $-i\omega \varepsilon_{Ag} E_0$, with $E_0$ being the local electric field in silver. In the limit of thin nanodipoles (i.e., $R \ll L$) an approximate standing wave displacement current distribution is expected along the dipole, with expression:

$$I(z) = I_0 \frac{\sin\left[\frac{\pi}{\lambda_{eff}}\left(L_{eff} - 2|z|\right)\right]}{\sin\left[\frac{\pi}{\lambda_{eff}} L_{eff}\right]}, \tag{1}$$

with $I_0$ being the current entering the nanoantenna arms at the feeding point. The corresponding optical radiation resistance may be evaluated in closed form (not reported here, since it is too lengthy, but plotted in Fig. 3) [16].

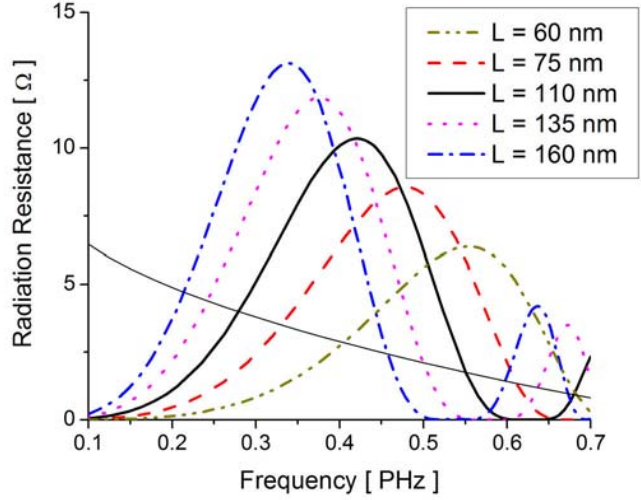

Figure 3 – (Color online) Radiation resistance for the nanodipoles of Fig. 2.

The predicted values of resistance in Fig. 3 are lower, but comparable in magnitude, than those expected from an RF dipole. This effect is due to the shortening of $\lambda_{eff}$ (compared with $\lambda_0$), which leads to an electrically smaller radiation aperture as compared with analogous classic RF dipoles. The thinner solid line in the figure refers to the theoretical radiation resistance evaluated for a nanodipole with $L_{eff} = \lambda_{eff}/2$ (at RF this line would be constant at $73\Omega$). This line crosses the different curves at their corresponding "short-circuit" resonance, validating the numerical results in Fig. 2. At this frequency, $I_{max} = I_0$, implying that the value of radiation resistance in Fig. 3 is "seen" at the input port. Due to the material losses in the nanoantenna, this value is somewhat lower, but comparable in magnitude, than the input resistance at the short-circuit resonant frequency, validating and explaining the numerical results for the radiation efficiency reported in Fig. 2c. The radiation efficiency may in general be given by

$$\eta_{eff} = \frac{R_{rad}}{R_{rad} + \sin^2 x_{eff} R_{loss}} = \frac{R_{rad}}{R_{in}},$$ where $R_{loss}$ is the portion of input resistance

associated with the power lost through absorption in the silver. $R_{rad}$ increases for longer antennas, as seen in Fig. 3, since longer nanodipoles resonate at lower frequencies, for which the effect of wavelength shortening is reduced. This is consistent with the corresponding increase in $\eta_{eff}$ predicted in Fig. 2c.

The above analysis of the nanodipole input impedance allows one to treat the two antenna arms as the "terminals" of a "lumped" element, which in turn would facilitate the design of a feeding waveguide, analogous to what is commonly practiced for conventional RF dipole antennas. This suggests the possibility of designing suitable "nanoloads" for this optical antenna using optical lumped circuit elements in order to adjust and tune the nanodipole resonant frequency or to "match" the nanoantenna with a feeding network at nanoscales. To this end, we "load" the nanodipole of Fig. 1 at its gap region with a nanoparticle (e.g. a nanodisk) made of a material with permittivity $\varepsilon_{load}$, that acts as an optical lumped element [i.e., a nanocapacitor (nanoinductor) when $\text{Re}(\varepsilon_{load})$ is positive (negative)]. In this scenario, the load effectively replaces the air gap and becomes in "parallel" with the antenna terminals, which can still be fed at the gap, for instance by a plasmonic waveguide, as sketched in the inset of Fig. 4 together with the modified circuit model.

Figure 4 reports the modification of the input resistance and reactance of the nanodipole, after being loaded with nanodisks of different permittivity. It is evident how the open-circuit resonant frequency may be tuned to the desired value by properly choosing the nanocapacitance or nanoinductance at the gap terminals, without changing the nanodipole length. We point out that this shift in the resonant frequency is not only predictable qualitatively, but it can also be quantified and tailored at will, similar to the antenna design at RF. Following our optical nanocircuit theory [14], the load impedance may indeed be easily evaluated as $Z_{load} = ig/(\omega \varepsilon_{load} \pi R^2)$, being capacitive (inductive) for positive (negative) $\text{Re}(\varepsilon_{load})$.

Since the nanoload is replacing the air gap, it is not surprising to discover that the open-circuit resonant frequency, as obtained in Fig. 4, *quantitatively* coincides with the frequency for which the value of the parallel load reactance $X_{load} = -\text{Im}(Z_{load})$ "goes into resonance" with the dipole's intrinsic reactance $X_{dip} = -\text{Im}[Z_{dip}]$ after proper de-embedding/removing the air gap impedance $Z_{gap}$. The different curves in Fig. 4 may indeed be obtained with very good approximation by considering the parallel combination of the input impedance of Fig. 2 with a capacitance represented by the "additional" parallel load $Z_{add} = (Z_{load}^{-1} - Z_{gap}^{-1})^{-1} = ig/[\omega(\varepsilon_{load} - \varepsilon_0)\pi R^2]$. For instance, when a silicon-nitride nanoload is used in place of the air gap, as seen in Fig. 4, the open-circuit resonance is shifted to $317\,THz$, at which $X_{in} = 700\,\Omega$ from Fig. 2b, and this is equal to $\text{Im}[Z_{add}]$ using the above expression. Similar considerations may be verified for the other loads considered in Fig. 4.

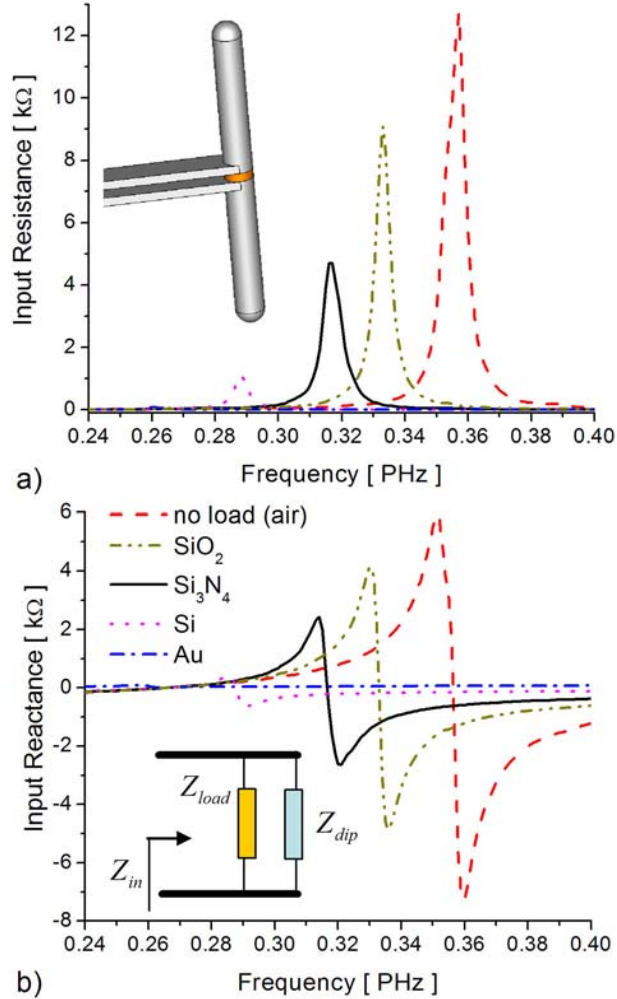

Figure 4 – (Color online) (a) Input resistance and (b) input reactance for the loaded nanodipole antenna of Fig. 1 with $L = 110\,nm$, varying the material of nanoload at its gap.

We also notice that since a larger value of permittivity introduces a larger capacitance for the nanoload, the open-circuit resonance shifts to a lower frequency. By contrast, using a nanodisk made of materials with $\text{Re}(\varepsilon_{load}) < 0$ (i.e., an inductive load), as it was done by considering a gold nanodisk (blue curve), shifts the resonant frequency downward to a value below the short-circuit frequency in the region where $X_{dip} < 0$. The "short-circuit" resonance is not modified by the presence of a parallel nanoload, since this resonance is due to the nulling of the intrinsic reactance of the nanodipole without the gap impedance. This resonance may, however, be tuned using a *series* nanoload, placed between the feeding terminals and the antenna arms (not shown here).

The above discussion effectively shows that the concept of nanoantenna loading is analogous to its RF counterpart, when the optical nanocircuit elements [14]-[15] are used to quantify the load impedance. An interesting corollary to this observation is that the input resistance at resonance is consistently reduced, when the permittivity of the parallel nanodisk load is increased (see Fig. 4a), again fully consistent with its circuit model. A judicious design of the parallel loads may allow a good matching between the nanodipole antenna and its feeding system at the open-circuit resonance (e.g., a plasmonic waveguide).

The results presented in this letter may provide an exciting method for tailoring the radiation performance and frequency design of optical nanoantennas and may facilitate the tuning and use of these nanoantennas for various applications in diverse fields such as optical communications, biological and medical sensors at the nanoscales, and nano-optical microscopy.

This work is supported in part by the U.S. Air Force Office of Scientific Research (AFOSR) grant number FA9550-05-1-0442.